# From Data Processes to Data Products:
## Knowledge Infrastructures in Astronomy

Christine L. Borgman\* (UCLA, christine.borgman@ucla.edu) and Morgan F. Wofford (University of Michigan, mwofford@umich.edu)

**Abstract**

We explore how astronomers take observational data from telescopes, process them into usable scientific data products, curate them for later use, and reuse data for further inquiry. Astronomers have invested heavily in knowledge infrastructures − robust networks of people, artifacts, and institutions that generate, share, and maintain specific knowledge about the human and natural worlds. Drawing upon a decade of interviews and ethnography, this article compares how three astronomy groups capture, process, and archive data, and for whom. The Sloan Digital Sky Survey is a mission with a dedicated telescope and instruments, while the Black Hole

Group and Integrative Astronomy Group (both pseudonyms) are university-based, investigator-led collaborations. Findings are organized into four themes: (1) how these projects develop and maintain their workflows; (2) how they capture and archive their data; (3) how they maintain and repair knowledge infrastructures; and (4) how they use and reuse data products over time. We found that astronomers encode their research methods in software known as pipelines. Algorithms help to point telescopes at targets, remove artifacts, calibrate instruments, and accomplish myriad validation tasks. Observations may be reprocessed many times to become new data products that serve new scientific purposes. Knowledge production in the form of scientific publications is the primary goal of these projects. They vary in incentives and resources to sustain access to their data products. We conclude that software pipelines are essential components of astronomical knowledge infrastructures, but are fragile, difficult to maintain and repair, and often invisible. Reusing data products is fundamental to the science of astronomy, whether or not those resources are made publicly available. We make recommendations for sustaining access to data products in scientific fields such as astronomy.

# 1  Introduction

Astronomy is an ideal domain in which to study how observations are processed to become useful scientific products, how data products are used and reused, and by whom. These data products may remain useful for decades or centuries. As a research community, astronomy has invested heavily in knowledge infrastructures, which are "robust networks of people, artifacts, and institutions that generate, share, and maintain specific knowledge about the human and natural worlds" (Edwards, 2010, p. 17). Observatories, telescopes, instruments, data archives, software tools, technical standards, and metadata services are among the many technical and institutional components of astronomical knowledge infrastructures (Ackerman et al., 2008; Burns et al., 2014; Genova, 2018; Genova et al., 2000; Gray, 2009; Hanisch et al., 2001; Hasan et al., 2000; Mink et al., 2014). Infrastructures are often invisible, learned through experience, and fragile, yet essential to the work of science (Star & Ruhleder, 1996). Like invisible astronomical objects, knowledge infrastructures are best studied indirectly by focusing on how they influence other phenomena.

This article builds upon previous work on scientific data practices conducted under grants to the UCLA Center for Knowledge Infrastructures (CKI) (Borgman, 2019; Borgman et al., 2020, 2021, 2016; Boscoe, 2019; Darch et al., 2021, 2017; Darch & Sands, 2015; Pasquetto et al., 2017, 2019; Sands, 2017; Scroggins et al., 2020; Scroggins & Boscoe, 2020; Scroggins & Pasquetto, 2020; Wofford et al., 2020). We draw upon a decade of interviews and ethnography to explore how astronomers take observational data from telescopes, process them into usable scientific data products, curate them for later use by themselves or others, and reuse data for various forms of inquiry. We compare three astronomy groups to examine how they employ knowledge infrastructures in their data practices; how those infrastructures are maintained and repaired, by whom, and when; and how their data products are used and reused. The article concludes with lessons learned about processing data into products that may be instructive for future research in astronomy and other sciences.

Astronomers are scientists who study celestial objects and other phenomena that originate beyond the earth's atmosphere. Observatories are among the most durable features of astronomical knowledge infrastructures, operating over the course of decades. These are institutions large or small, space-based or ground-based. "Observatory" usually refers to the overall mission, such as Hubble or Spitzer, or sometimes to a geographic location with multiple missions, such as the Mauna Kea Observatories. The Chandra X-Ray Observatory, for example, includes a spacecraft launched in 1999 with four pairs of mirrors and four scientific instruments consisting of cameras and spectrometers. The observatory mission also includes operations on the ground to support data collection, software development, pipeline processing, the Chandra Data Archive, and other services (*Chandra Data Archive*, 2019; *Chandra Science Instruments*, 2021). Similarly, other space-based observatories such as Hubble, Spitzer, and James Webb, and ground-based observatories such as Mt. Wilson Observatory in California and the Mauna Kea Observatories in Hawaii, may support multiple telescopes and instruments that can be configured to acquire specific images or spectra of astrophysical phenomena. Observatory staff manage facilities, instrumentation, and data with varying degrees of involvement from the astronomers who acquire data from these sites (Hoeppe, 2014, 2018, 2020b).

Between the 1960s nd the 1990s, astronomy made the transition from analog to digital technologies, a process that led to fundamental changes in astronomical practice (McCray, 2004, 2014). As a community, astronomers have developed a far more comprehensive and coherent infrastructure than have other scientific domains (Borgman et al., 2016; Genova et al., 2017). One reason for this coherence is the agreement on "a single sky" (Hoeppe, 2019; Hogg, 2014; Munns, 2012). By using the sky as an organizing principle of their science, the community can agree upon a coordinate system and associated metadata standards such as FITS files (Accomazzi et al., 2014, 2015; Committee on NASA Astronomy Science Centers, & National Research Council, 2007; Hanisch et al., 2001; Scroggins & Boscoe, 2020; Vertesi, 2015, 2020).

Layered over these agreements are a sophisticated array of value-added data services such as the International Virtual Observatory; the Strasbourg Astronomical Data Center – best known by its French acronym CDS – which hosts data and metadata systems such as SIMBAD, VizieR, and Aladin; IPAC at Caltech, originally known as the Infrared Processing and Analysis Center, which hosts NASA, Jet Propulsion Lab (JPL), and NSF-funded databases; the Space Telescope Science Institute, operated by the Association of Universities for Research in Astronomy, which hosts archives of the Hubble Space Telescope and numerous other space missions; and the NASA Astrophysics Data System, which is a comprehensive bibliographic database for the domain (Centre de Données astronomiques de Strasbourg, 2021; Genova, 2018; *Infrared Processing and Analysis Center*, 2021; *Space Telescope Science Institute*, 2021; NASA/ADS, 2021). Astronomers rely on these knowledge infrastructure components in their daily work. Many components emerge from practice, as systems and services evolve over time. As Susan Leigh Star (1999, p. 382) noted in an early paper on the topic, "nobody is really in charge of infrastructure."

## 2   Theories and Themes

This article emerges from a long-term research agenda to explore the origins of scientific data, how they evolve through processing and analysis, how they are used and reused over time, and by whom (Borgman, 2015; Borgman et al., 2012; Pasquetto et al., 2017, 2019; Wallis et al., 2013). Sabina Leonelli and collaborators similarly have found that tracing "data journeys" is a fruitful approach to exploring scientific practice (Leonelli, 2010, 2020; Leonelli & Tempini, 2020). "Data" remains the most contested concept in data science, as discussed in the first issue of this journal (Borgman, 2019; Leonelli, 2019; Wing, 2019). Rather than provide an extensive explication here, we simply submit that observations become scientific data when used as evidence of phenomena (Borgman, 2015; Leonelli, 2016).

The three astronomy projects used as exemplars here have substantially different workflows, processing practices, and ways to maintain their data and software. They also differ in their relationship to their observatories, their scientific questions, their uses of the resulting datasets, and their reliance on the knowledge infrastructures of the astronomy community. The Sloan Digital Sky Survey is a mission with a dedicated telescope and instruments, while the Black Hole Group and Integrative Astronomy Group (both are pseudonyms) are investigator-led collaborations based at different universities. Relationships between sky surveys and investigator-led projects are synergistic. Sky surveys observe particular parts of the sky systematically, documenting certain phenomena in depth and over time. Investigator-led groups often follow up on interesting observations from surveys to determine what is occurring and why, and may combine data products from multiple sources. Conversely, findings from studies initiated by small investigator-led groups may contribute to the design of future sky surveys.

In considering how astronomical observations become useful scientific products, some clarifying terminology is necessary. We use "data product" as the scientifically useful entity resulting from software processing of observations, and "data production" as those processing activities. Baker and Mayernik (2020) explain data production in the narrower sense of producing data "with the intention that they may be used by others" (Baker & Mayernik, 2020, p. 3), work that is usually the responsibility of observatories and archives. The latter subset of data production activities includes assembly, packaging, adding metadata, contributing to repositories, archiving, and maintaining access to those data. The authors make a further distinction between data production and knowledge production, using examples from the earth and space sciences. They found that scientists focus primarily on knowledge production, which includes collecting, processing, cleaning, and interpreting data for the purposes of writing journal articles and other scholarly reports.

Baker and Mayernik also found that scientists invest in their data as evidence for knowledge production, but few perform the additional work necessary to make their data products available as a community resource. Scientific stakeholders, including observatories, investigators, and funders, balance their investments in knowledge and data production work. These choices influence the sustainability of data resources. The distinction between producing data products for a research team's own use or for public distribution provides a theoretical framework for this article. A central goal of open science is to release and share data in ways that those data become Findable, Accessible, Interoperable, and Reusable (FAIR) (Wilkinson et al., 2016). The FAIR principles are a very high bar, and one rarely met in day-to-day scientific practice (Borgman, 2015). While we do not endeavor to test the FAIR principles directly, we do examine what data are being produced, to what extent those data are reusable by others, and who may be able to use those data in the future.

Rather than explain our findings about each astronomy project individually, we organize our results by four themes, after a brief foray to explain our research methods. The first theme compares how these projects develop and maintain their workflows, and ways in which their research methods are embedded in their data processing practices. Concepts of data processing and data products are developed in theme 1, as are the ways in which these activities depend on other knowledge infrastructure components. The second theme explores details of how these astronomy projects capture and archive their data, identifying factors by which these practices differ. The third theme considers the robustness and fragility of knowledge infrastructure components, examining the kinds of maintenance and repair required, by whom these activities are performed, and at what stages of data processing. The fourth and last theme examines the uses and reuses of astronomy data products over time, identifying stages of activity, and addressing questions of audiences and utility of scientific data products.

## 3   Research Methods

Findings reported here are based on our studies of astronomers and astronomy research staff conducted under a series of grants to the UCLA Center for Knowledge Infrastructures and its predecessor teams. The first of these endeavors began in 2008 with the Sloan Digital Sky Survey, evolving through several grant projects with foci on knowledge transfer, data management, and workforce development (Borgman et al., 2015; Darch et al., 2020, 2021; Sands, 2017). Research with the Integrative Astronomy Group, a pseudonym assigned for confidentiality, started in 2009 to study how individual astronomy research groups manage multiple sources of data (Borgman, 2015; Borgman et al., 2016). Our partnership with the Black Hole Group and Vida Observatory, also pseudonyms, began in 2016 with a focus on learning how astronomers keep data alive over decades (Boscoe, 2019).

We employ semi-structured and unstructured interviews, ethnographic participant-observation, and document analyses. Our mixed methods enable us to examine data practices over extended time frames and between sites. Interviews, typically conducted at the participants' worksites, usually last between 45 and 120 minutes. We record interviews (except in the few cases where participants asked not to be recorded), have the recordings transcribed by a professional service, correct the transcriptions for scientific terminology and other errors, and then code the transcripts using qualitative analysis tools. Table 1 summarizes the set of interviews, people, institutions, and time periods used in this study.

**Table 1: Astronomy interviews analyzed for this article**

| Sites | Interviews | People | Institutions | Period |
| --- | --- | --- | --- | --- |
| Sloan Digital Sky Survey | 149 | 131 | 35 | 2009-2018 |
| Black Hole Group | 17 | 13 | 3 | 2016-2020 |
| Integrative Astronomy Group | 22 | 8 | 1 | 2009-2020 |
| Total | 188 | 152 | 39 | 2009-2020 |

We conducted ethnographic participant-observation at each of these sites over the same period. Notes from these observations inform this article, which otherwise draws on the interview transcripts and recordings. Also included as background materials are public and private documents associated with participants at these sites. In the analyses reported here, we used NVivo 11, a qualitative analysis software package, to code interview transcripts, field notes, and documents (*NVivo 11*, 2015). We applied grounded theory approaches to these analyses (Glaser & Strauss, 1967). More details are provided in a recent methodology article covering 20 years of research by the CKI (Borgman et al., 2021).

To update and verify our findings for this article, we invited members of each of the three astronomy projects to comment on earlier drafts. We revised the manuscript based on the substantive comments received.

## 4  Theme 1: Workflows, Methods, and Data Processes

Data is a fluid concept in astronomy, exemplifying the notion that "raw data is an oxymoron" (Gitelman, 2013). Generally speaking, observations flow as digital signals from telescopic instruments to processing sofware, and from there to the local data stores of scientific research teams and to data archives associated with the observatory. Processing software, which astronomers refer to as "pipelines," consists of algorithms to transform "raw data into valuable information for the intended audience via data reduction and analysis" (Cabral et al., 2017, p. 140). Pipelines are far less linear than the term implies, however.

The design of new astronomical instruments is based on findings from prior instruments, on simulated data, on theoretical models, and on scientific conjectures about what could be found in certain wavelengths, with certain capabilities, and certain functions. Instruments on ground-based telescopes may be upgraded repeatedly, extending the life of observatories. Each new dataset is a consequence of earlier scientific decisions, hence the origin of astronomy data can be an infinite regress (Borgman, 2015). Boscoe (2019) provides an extensive discussion and informative diagrams of how astronomy data travel through workflows and pipelines across public and private boundaries of astronomical knowledge infrastructures.

Data processing algorithms are the "secret sauce" of astronomers. These algorithms, or pipelines, were frequently mentioned as "our methodology" or "our recipe." Where processing observations begins and ends depends upon one's vantage point. At the observatory instrument, the "raw data" may be the photons, which become electrons when they hit the detector, and become processed data after minimal corrections and addition of metadata. That observatory dataset may be the "raw data" that becomes input to software pipelines built by an investigator's team. Similarly, data reduction to remove artifacts is sometimes viewed as "pre-processing" and at other times as "post-processing" (Levesque, 2020).

Calibration, which is the act of comparing a measurement to a standard of known accuracy, is fundamental to scientific data processes (Plant & Hanisch, 2020). In a home kitchen, comparing the temperature on an oven display to the reading on an inexpensive thermometer placed in the oven may suffice for accuracy. In astronomy, far more precision is required. Calibration involves comparisons to multiple measurements of stellar objects, phenomena, instrument and weather conditions, simulations, and other entities specific to the scientific problem under study (Hoeppe, 2014, 2020a). Astronomers conduct numerous validity checks to remove sources of error or artifacts that may influence scientific interpretation (McNutt, 2020; Plant & Hanisch, 2020). Different kinds of instrument calibration may occur before, during, and after an observing run. Similarly, observations may be calibrated against known results from other instruments at other times.

Astronomers build processing software at many different levels of granularity. Each instrument in an observatory may have a dedicated pipeline. For example, at the time of commissioning in 2005, investigators delivered a custom data reduction pipeline with OSIRIS, which is a near-infrared integral field spectrograph at Keck Observatory in Hawaii. After a decade of observations, astronomers made major improvements to the pipeline and calibrations, reporting on lessons learned (Lockhart et al., 2019). Astronomers simulate pipelines as part of designing new instruments, such as the Thirty Meter Telescope (Wright et al., 2016). Investigators also construct their own local pipelines to process observations for specific science problems and experiments.

A point of contention in our analyses was how to distinguish between the overall workflow of data handling and specific stages of processing observations into scientific data products. Generally, the knowledge production stages of the astronomy teams studied are to (1) design a study of astrophysical phenomena, (2) point a telescope with configured instruments at a science target, (3) take observational data, such as images and spectroscopy, (4) process observational data through software pipelines, (5) analyze the reduced data to produce scientific findings, and (6) write papers, usually accompanied by datasets that constitute evidence for the findings. Data production stages of observatory and archive teams include activities to curate, archive, maintain, and make available the resulting data. For the purposes of this article, astronomy observational workflows comprise all of these stages.

In the ideal case, all hardware and software function effectively on observing nights, the necessary staff are available to manage the workflows, and the seeing conditions are good to excellent. Much can go right, and much can go wrong along the way.

Figure 1 represents the common activities of astronomy data workflows across the three projects studied, focusing on pipeline processes in stages 2 through 5 and disposition of data products. Each observatory has a staff of astronomers, instrument specialists, technical specialists, and other people who support the operations. These individuals have primary responsibility for operating, maintaining, and repairing the telescope, instruments, and facilities. Observations typically flow from telescopic instruments through software pipelines both to the research groups who are taking experimental data and to data archives associated with the observatory.

**Figure 1: Overview of an astronomy data processing workflow**

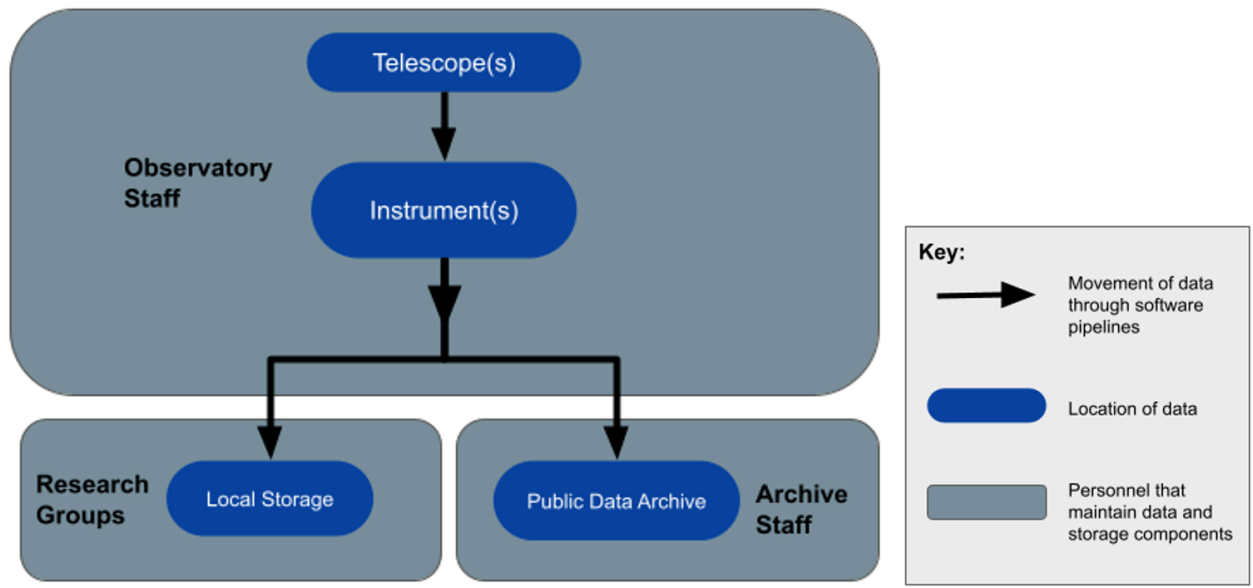

### 4.1 Sloan Digital Sky Survey

The Sloan Digital Sky Survey (SDSS) began survey observations in 2000, after investing more than a decade in design and construction. SDSS is a pioneer in pipeline processing, cosmology, modern observational survey methods, and scientific data archiving (Finkbeiner, 2010; Bell et al., 2009; Hey et al., 2009; Szalay, Gray, Thakar, Kunszt, Malik, Raddick, Stoughton, et al., 2002). The primary telescopes for SDSS are located in Apache Point, New Mexico (U.S.), part of the Apache Point Observatory operated by the Astrophysical Research Consortium. Instruments in Chile provide additional observations of the Southern Hemisphere. The survey covers one-quarter of the night sky to provide high quality optical and spectroscopic imaging data on galaxies, quasars, and stars. Four phases of the project are now complete, with a fifth phase under way from 2020 to 2025. The first phase, SDSS-I, spanned 2000 to 2005; SDSS-II, 2005 to 2008; SDSS-III, 2008 to 2014; and SDSS-IV, 2014-2021. Over three decades from the proposal to current operations, the survey has evolved to incorporate new funding sources, partners, leadership, technologies, and scientific goals (*Apache Point Observatory*, 2021; *Astrophysical Research Consortium*, 2021). While SDSS has maintained continuity of the site and telescopes, each of the five surveys has taken the science in new directions. SDSS-V, for example is entitled *Mapping the Universe*, which endeavors to pioneer panoptic spectroscopy (Sloan Digital Sky Survey, 2021).

Our interviews and ethnography focused on the first two phases of the SDSS project, with some later follow-up (Borgman et al., 2016; Darch et al., 2020, 2021; Sands, 2017). During SDSS I and II, the two telescopes at Apache Point were taking all the observations. The Sloan Foundation 2.5-Meter Telescope is the survey's main telescope; the NMSU 1-Meter Telescope observes stars too bright for the larger telescope to see. Multiple instruments on the 2.5-Meter Telescope, as shown in Figure 2, collect optical images and spectroscopy. Observations flow through data pipelines from the 2.5-Meter Telescope to the SDSS data archive; data users can download datasets from the public data archive to their local storage.

Observing staff at Apache Point maintain the physical facility, telescope, and instruments. Pipeline processing, which constituted about 25% of the cost of the initial project (Szalay et al., 2001), was divided into multiple modules and parceled out to SDSS partners at Fermilab, University of Chicago, University of Washington, Johns Hopkins University, Princeton University, and elsewhere. SDSS is known for its innovative pipeline processing and search software, built in partnership with Microsoft and the legendary computer scientist, Jim Gray. SDSS I/II, which we studied most closely, produced open data using proprietary software (Darch & Sands, 2017; Gray et al., 2005; Szalay, 2008; Thakar et al., 2008).

**Figure 2: Overview of SDSS data processing workflow**

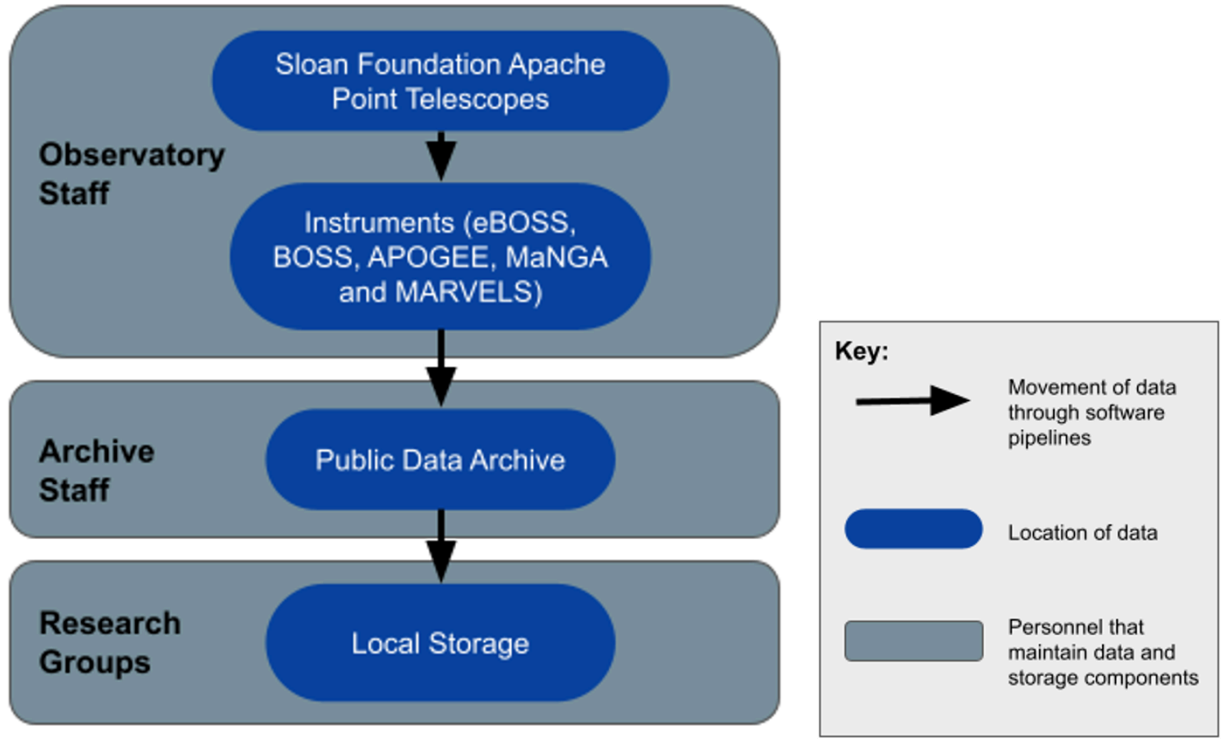

SDSS relies on the knowledge infrastructures of astronomy in numerous respects. They calibrate their instruments against star catalogs and other measurements from prior studies, and sometimes compare them to other telescopes, such as the 3.5 Meter Telescope at Apache Point. They produce datasets as FITS files, using common coordinate systems. SDSS also contributes to the KI of astronomy by developing new pipelines, producing and maintaining open data archives, and releasing their data to value-added services.

## 4.2 Black Hole Group

The Black Hole Group (BHG) is a university-based research team with about 25 astronomers, including students and research fellows. They have partners at universities and observatories in the U.S. and Europe. BHG began in the 1990s, and some original team members are still involved in the group. Their "key science problem" is to understand the physics of black holes and their environs, and complementary phenomena such as nearby gas clouds and star formation. Temporal studies, known as "time-domain astronomy," are unusual and complex because they require accumulating consistent data products over long periods of time. This team combines new observations with data taken over the course of about 25 years. Instruments, storage, and computation have changed radically over this time period, as has their understanding of the science. They are pioneers in data collection methods and in analytical methods to combine images and spectroscopy into other data products such as orbits of stars near black holes.

The Black Hole Group has a synergistic relationship with Vida Observatory, thus we examine their independent and combined approaches to processing observations into data products. The Vida Observatory operates large ground-based telescopes tethered to a suite of instruments that support observational astronomers from various subdisciplines of astronomy and astrophysics.

BHG collects most of their data from Vida and thus has vested interests in Vida's infrastructure. The group has collaborated with Vida to develop, deploy, and maintain several of the observatory's instruments and associated pipelines. Vida astronomers, in turn, collaborate on some of the BHG science projects. However, the partnership is not exclusive. BHG collects data from multiple observatories and has many other collaborators. Astronomers from around the world submit competitive proposals to collect data at the Vida Observatory, hence the observatory serves a large community with diverse scientific goals. Many members of the community participate in pipeline development and maintenance.

To simplify the complex workflows of the Black Hole Group, we restrict our analysis to how the BHG acquires and processes data from the Vida observatory, the disposition of these data products, and their relationships to knowledge infrastructures. The initial data processing workflow originates with configured instruments placed on a Vida telescope and ends at both the Vida Observatory Archive and BHG local storage. Figure 3 is a high level overview of these workflows, drawn from interviews with both groups.

**Figure 3: Overview of data processing workflows from Vida Telescopes**

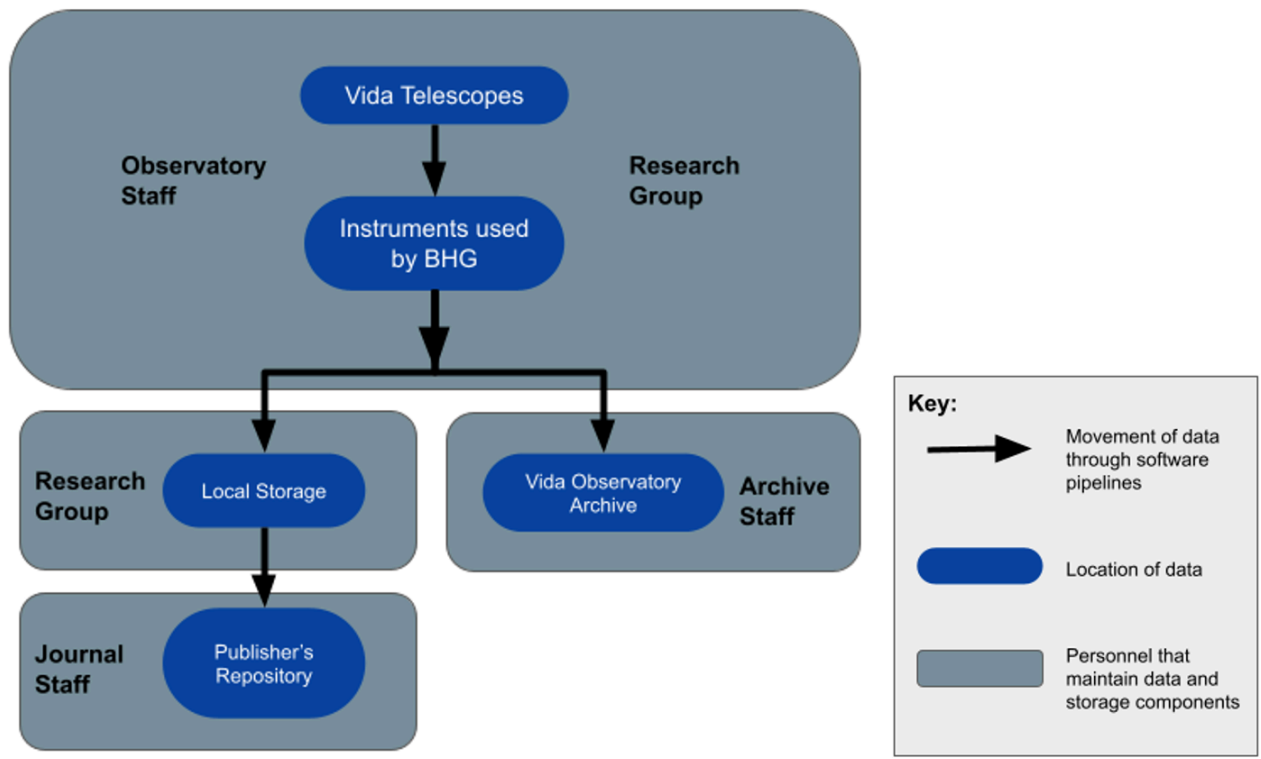

BHG astronomers distinguish between three types of pipelines used by Vida, based on who is responsible for developing and maintaining them. Vida is solely responsible for instrument pipelines that are direct interfaces to instruments, "such as those that add FITS header keywords to the files read out from the detectors" (BHG astronomer, 2020). A second category is instrument pipelines jointly maintained by Vida staff and astronomers in the community, such as a specific spectrograph pipeline. A third category is pipelines maintained by the Vida Observatory Archive to process data into common formats for observations from all the Vida instruments. All of these pipelines can be viewed as components of the knowledge infrastructure of the observatory.

Once BHG receives a dataset from a night's observing, after initial processing by Vida, BHG researchers process those data through their own pipelines for further calibration and analyses. These are complex pipelines that branch in many directions based on types of data and on science questions. The team continually adapts their pipelines for changes in instruments, new data types, and new science questions. They also maintain basic functionality in their suite of pipelines so that they can make comparisons over time.

The Black Hole Group benefits from astronomical knowledge infrastructure investments such as the technologies and pipelines of the Vida Observatory, standards such as FITS files and coordinate systems, and the ability to compare their measurements to known sources from the Hubble Space Telescope, SDSS, Gaia, and other observatories. BHG contributes to the KI of astronomy through participation in building Vida components, developing generalizable methodologies, and other means. However, because much of their time-domain science focuses on faint stars not previously known, they are less able to use star catalogs and related value-added data services, as discussed in Theme 2.

### 4.3 Integrative Astronomy Group: Star-Forming Region Survey

The Integrative Astronomy Group (IAG; a pseudonym) is an open science team that brings together astronomers, computer scientists, information scientists, and librarians to develop tools and infrastructure for data-intensive astronomy. They have no astronomical agenda of their own; rather, they facilitate scientific research through innovations in infrastructure. Their informal collaboration began in the early 2000s, developing tools that are widely adopted in astronomy practice. The expressed mission of the IAG is to make infrastructure that anyone can use.

The Star-Forming Region Survey (SFRS) is a pseudonym for a multi-year collaborative project led by the same astronomer who directs the Integrative Astronomy Group. The collaboration was formed to address a specific set of scientific goals, bringing together an international group of astronomers with complementary expertise. This survey also began in the early 2000s, marking a foray into open tools, integration of old and new data, and release of open data to the astronomy community. The project compared the physics of three known star-forming regions and molecular clouds, using data from multiple instruments taken at various wavelengths.

As shown in Figure 4, the Star-Forming Region Survey, conducted in partnership with the Integrative Astronomy Group, is comprised of data from six sources: three public data archives and three sets of observations acquired by the investigators from new proposals. They seeded the study with data from the public archives of two surveys, one originating at a space-based observatory (S1) and one at a ground-based observatory (G1). After initial analysis of these data products, they submitted observing proposals to three ground-based observatories (G2, G3, G4) to acquire complementary data on the same region. The public data archive associated with one of these three observatories (G2) also yielded data products that were included in the SFRS.

**Figure 4: Overview of data processing workflows to create the Star-Forming Region Survey**

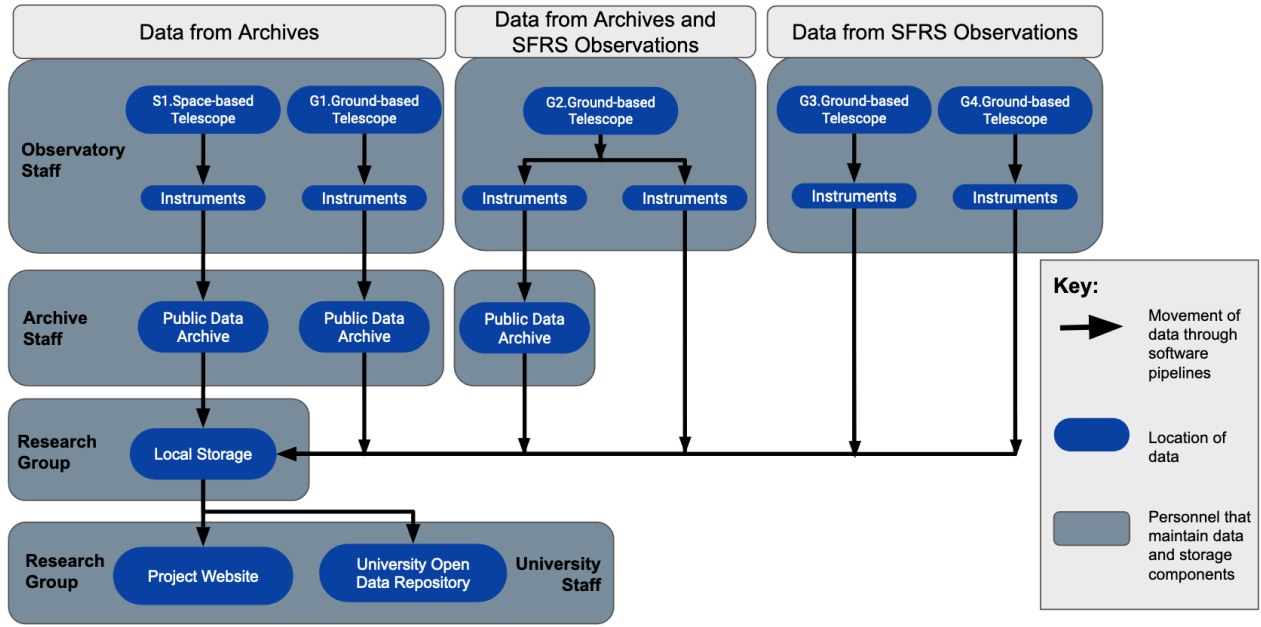

The SFRS collaboration employed IAG tools and new code to assemble the survey dataset. The observatories from which they obtained data performed initial processing. Calibrations done by the ground-based observatories were sufficient for SFRS purposes; the team reprocessed datasets acquired from the space-based observatory archive to correct calibration errors. Once they completed the multi-year process of data acquisition, the SFRS collaboration wrote pipelines to integrate those data products into a unified dataset in local storage. Analyses of that dataset resulted in multiple publications by the investigators. The SFRS data products were posted on the project website and submitted to the open data repository of the PI's university, as shown in the bottom box of Figure 4.

Among the components of astronomical knowledge infrastructures on which the survey depended were telescopic instruments, data archives, software pipelines, technical standards, and coordinate systems that facilitate data integration. They employed the Integrative Astronomy Group's tools and methods to integrate heterogeneous data products. IAG tools were useful in extracting header information from FITS files, in combining and visualizing data, and in new forms of digital publishing.

### 4.4 Theme 1 Discussion

These three projects began in the 1990s and early 2000s with similar access to data standards, technologies, software pipelines, archives, and other astronomical knowledge infrastructure. Sophisticated code transforms "raw data" into useful scientific products. In the process of data reduction, these pipelines remove artifacts, calibrate observations against known metrics, and perform myriad other validation checks. How these astronomers enact their research methodologies in software varies considerably by their scientific goals.

Our four figures that represent workflows highlight the multiple pipelines and actors involved, while greatly simplifying processes that are always in flux. Scientists adapt workflows to new science questions, technologies, partners, and observing conditions. They find workarounds for malfunctioning equipment or unavailable staff. A full map of one of these worklows would be a snapshot in time that suggests a dynamic network of intersecting pipelines with multiple branching points.

In the years of our studies, these projects both contributed to, and benefitted from, astronomical knowledge infrastructures. They differ in workflows, research methods, and data processing, and in how they rely on astronomical knowledge infrastructure components such as observatories, telescopes, instrumentation, data archives, journals, funding agencies, technical standards, software tools, conferences, and professional societies that serve the global astronomy community.

## 5 Theme 2: Capturing and Archiving Data

Workflows and pipelines are a means to an end: capturing data that become valid scientific evidence. Astronomers design experiments that require observations from certain instruments, at certain times, under certain conditions. They write observing proposals to obtain the desired data. They may write other grant proposals to fund their research groups for knowledge production labor such as developing the science, taking the data, writing local pipeline code, analyzing those data, writing the papers, and maintaining those data for subsequent knowledge production. Astronomers also design studies to acquire observational data from one or more archives. They may process old or new data through existing pipelines or write new pipelines to address their scientific problems (Hoeppe, 2020a; Zucker et al., 2019).

Modern astronomical observatories provide their investigators with data products in calibrated, canonical digital formats. They also process and package these data for deposit into dedicated archives. Examples include the Sloan Digital Sky Surveys, Hubble Legacy Archive, and Chandra Source Catalogs. Investigators typically retain control over their observations for a proprietary period before public access. Datasets in these archives are embargoed for those agreed time periods.

In Theme 2, we look more closely at the steps involved in capturing and archiving data, which demonstrates how "raw data" from a night's observing may diverge into multiple, and malleable, data products over time.

### 5.1 Sloan Digital Sky Survey: Capturing and Archiving Data

Capturing reliable data from the SDSS observatory involves a wide array of hardware, software, and human expertise. Developing the software to control the telescope and its suites of instruments, monitor interconnections among systems, and manage data flow were major investments of the SDSS collaboration. Daily operations at Apache Point are managed by observing staff, with night-time management by astronomers who supervise data capture.

The SDSS collaboration devoted many years to determining science targets and the precision with which they could be measured. Ann Finkbeiner's book (2010), *A Grand and Bold Thing,* tells the SDSS story in great detail, complementing and reinforcing findings from our ethnography and interviews.

Observational data are captured nightly at Apache Point, to the extent that weather and instrument conditions allow, except during the summer monsoon season. SDSS acquires both images and spectroscopic data. Cameras acquire images, which are processed through photometric pipelines. In the early stages of SDSS, these images were used to identify targets for spectroscopy. Those spectroscopic targets were then drilled into aluminum plates, 640 holes per plate, each with a cable to the spectrograph. Multiple plates might be swapped out per night.

Despite being a sky survey, the data collection process is far from robotic. On any given night, astronomers might acquire images, spectroscopy, or both. These choices are based on numerous parameters that include overall scientific goals for the observing season, for specific nights, weather conditions suitable for images or spectroscopy, and equipment function.

Other pipelines were written for spectrometry, for mapping the fibers in the disks to galaxy observations, for removing dust from images, and so on. Pipelines were rewritten to improve speed and accuracy, and to revise scientific goals. By breaking new scientific ground in data processing, SDSS was on the "bleeding edge" of technology. Tensions sometimes arose between partners over control and responsibility for these many interdependent software components (Finkbeiner, 2010).

The Sloan Digital Sky Survey also pioneered open data release, creating high quality archives for use by the community. The scientific investigators who proposed, designed, and manage the SDSS mission are entitled to control the data for a short proprietary period in advance of general release. SDSS has prioritized timely data releases over proprietary periods, sometimes forfeiting part of their allowable control period when they encountered excessive processing problems (Sands, 2017).

SDSS releases datasets only about once per year, due to the labor involved in processing, validating, and writing scientific documentation. The most recent data release (DR16) includes optical spectra of galaxies, stars, and quasars; infrared spectra; data cubes and maps; other spectra and imaging from prior SDSS programs; and value-added catalogs created throughout the first four phases of the mission (*Data Release 16 | SDSS*, 2020). These are cumulative datasets, each with extensive scientific documentation, some of it published as peer-reviewed journal articles to accompany individual data releases. The article associated with the Ninth Data Release, for example, has 236 authors, thus giving credit to the scientific and technical staff responsible for creating and maintaining the dataset (Ahn et al., 2012). These articles provide great detail about metadata, calibration, and pipeline parameters. Collectively, observations from the SDSS mission constitute a historically significant astronomy dataset in terms of its scope, quality, access, and extent of users (Finkbeiner, 2010; Plant & Hanisch, 2020; SDSS Collaboration, 2014, 2015, 2020a, 2020b, 2020c; Szalay, Gray, Thakar, Kunszt, Malik, Raddick, & Stoughton, 2002).

## 5.2 Black Hole Group and Vida Observatory: Capturing and Archiving Data

The Black Hole Group, like other astronomy investigators, typically is granted time on the Vida telescopes in quantities of quarter nights, half nights, or full nights. The number of hours constituting an observing night varies by time of year; winter nights are longer and summer nights are shorter. Investigators' proposals for telescope time request specific instruments in specific date ranges and at times of night when their science targets are visible from the Vida Observatory, justified by the science to be accomplished. Telescope time is a scarce and expensive resource, hence the proposal process is highly competitive. BHG writes multiple proposals per year for telescope time and for funding to support the research team.

On observing nights, BHG astronomers are paired with Vida staff who are experts on the instruments to be deployed. Vida support astronomers tend to specialize in one or two instruments, which enables them to develop necessary knowledge for night support, training observers, preventive maintenance, repairs, upgrades, and other kinds of collaboration with investigators. Vida staff run a series of tests and calibrations prior to the start of each observing run.

BHG astronomers also take their own measurements before, during, and after their observing runs to calibrate their data. For example, prior to observing, they usually take "darks," which are dark current measurements on the detector while the dome is closed. They may also take "flats" on a screen of known characteristics inside the dome. Once observing begins, they identify their first scientific target on the sky by coordinates, then use known nearby stars to isolate a field of view. One camera is used to acquire a target, then another instrument "locks on" to the target, after which the telescope can track the target across the sky as the earth moves. Once locked on and tracking, BHG astronomers and their Vida partners can obtain images or spectra with their configured instruments. While each observing run is carefully planned to capture certain science targets with certain instruments and parameters, BHG astronomers have backup plans to ensure they can make the best possible use of their precious time on the telescope.

Weather is always an issue for ground-based telescopes. BHG astronomers track weather conditions very closely in advance of an observing run and during the run, adapting in real time to seeing conditions. If seeing is inadequate to acquire quality data on their primary scientific target, then they will collect data for secondary or tertiary targets that will advance their overall scientific goals. Astronomers must also contend with events that can trigger the shutdown of instruments, which then must be calibrated to start up again. These events include bad weather, earthquakes, hurricanes, floods, volcanic eruptions, equipment malfunctions, and scheduled stops required by military security, ranging from a few seconds to a few minutes. Occasionally, bad weather or equipment problems prevent the dome from being opened all night. The BHG team makes their observing plans with an expectation of losing about one-third of their time, on an annual basis, to issues such as these.

Vida Observatory data flow both to a dedicated archive and to observers, resulting in multiple datasets with different characteristics and different states of processing. The Vida Observatory Archive contains data processed through instrument pipelines and then merged into a common format. These reformatted data products are available to the public when the proprietary period granted to observers has ended.

BHG astronomers can download their datasets shortly after they complete a night of observing, once the Vida team has processed the data through their pipelines. The BHG team processes their FITS image files and FITS spectroscopy files separately, as each has its own pipeline. They employ a combination of published software and locally written code, which together comprise their scientific methods for analyzing these data. Many of their grants are for software development. They reprocess data repeatedly to combine data from multiple nights in observing runs, to compare data from different epochs, and to integrate legacy data into new pipelines. BHG's code for photometry, astrometry, and spectroscopy also enables them to combine multiple data types to calculate orbits and astrophysical phenomena such as gas clouds.

Among the many data products that result from BHG pipeline processing are star lists. These are star catalogs specific to the region observed over the roughly 25 years of their time-domain studies. BHG calibrates their observations against Hubble, Gaia, SDSS, and other sources wherever possible, but few of their target stars are included in the value-added data services such as SIMBAD. As a consequence, BHG is less able to rely on astronomical knowledge infrastructures than many other projects. In comparing their star lists to those of other teams, they find inconsistencies in labeling, even in charts published by the same authors at different times. Metadata management in their local data archive is a continuing challenge.

### 5.3 Star-Forming Region Survey Data Capture and Archiving

Astronomers collaborating on the Star-Forming Region Survey collectively had expertise on the many instruments and technologies involved in acquiring these data. As the principal investigator of IAG stated in a 2010 interview: "Your…average scientist or astronomer, who's not building instruments…just wants the data. [They've] done multi-wavelength science…with radio data, optical data, infrared data, X-ray data, etc." No single astronomer is an expert at all the instruments involved in studies as complex as this one.

A first step in the survey was to identify existing data for the three star-forming regions of interest. Data archives for two ground-based telescopes and one space-based telescope held relevant observations. They searched these archives using coordinate-based and object-name-based searches. The archive of ground-based data yielded near-infrared extinction maps for the star-forming regions. The space-based astronomy archive was used to determine flux densities of the star-forming regions. While these results formed the core of their dataset, the team acquired about four times more data from three ground-based telescopes. To obtain these new data, the SFRS collaboration submitted observing proposals to three ground-based observatories. The team also developed pipelines to combine all of these data products into thermal dust emission maps.

Given the early days when the Star-Forming Region Survey was accomplished, one of their most significant contributions was to archive these data in ways that remain useful to the community. The Survey collaboration posted "all datasets immediately and in their entirety. [They] did not wait until papers were published or tie datasets only directly to papers," per the principal investigator (follow-up interview in 2020). The new data they collected (about 80% of the dataset from the three star-forming regions) and the data obtained from public archives continue to be reused by the astronomy community. The Survey collaboration also posted datasets "tied 1:1 with papers." These datasets remain available in university repositories, in publishers' warehouses, on investigators' websites, and elsewhere.

### 5.4 Theme 2 Discussion

All three of these projects follow the same basic practices to capture and archive data, as outlined in Figure 1. Investigators and their teams plan their observing nights carefully, well in advance, with contingency plans to accommodate seeing conditions. Observatories conduct initial data processing to clean data, remove artifacts, and release them in standardized formats. Investigators then process those data products further, usually using their own custom-designed pipeline software, to create new data products.

The specifics of data capture and archiving differ considerably by scientific goal. SDSS, as a multi-decade sky survey, has the largest scale of these projects. Their data production is intended for use by others, and is curated and maintained accordingly. The SDSS archive has extensive search capabilities, documentation, and professional assistance to exploit these data products. Both incremental and cumulative data releases remain available.

The Black Hole Group, in contrast, is a small research team. They are neither observatory nor archive nor survey. They invest in local data stores and software for their own long-term use. Vida Observatory, their partner, releases investigators' data on a periodic basis as proprietary periods end. Vida datasets are highly heterogeneous, reflecting the diversity of scientific targets and research methods employed by the observers in the community.

The Integrated Astronomy Group, a continuing collaboration focused on tool development, and the Star-Forming Region Survey, constitute one grant project with a narrow goal of creating a specific set of data products. They wrote their own software pipelines to integrate data from public archives and from their own observations. They released the dataset, along with their own publications and documentation, and moved on to other projects. The dataset remains available because the principal investigator's university continues to host it.

# 6 Theme 3: Maintaining and Repairing Knowledge Infrastructures

Software pipelines, digital data, and data archives are all inherently fragile due to continual upgrades of hardware and software and delicate interoperability between components. Over time, hardware can become less reliable, software can become incompatible with other tools, and bits can rot for lack of refreshing. Software documentation, especially open-source software popular in scientific applications, is difficult to maintain (Geiger et al., 2018). Dependencies abound. "Software collapse" (Hinsen, 2019) is a particular problem in these layered systems. When one component, such as part of an operating system, programming language, or analysis tool, ceases to function, every layer above it in the stack may collapse.

Maintenance is the work that keeps knowledge infrastructures running. It includes acts of supervision, care, repair, restoration, and the articulation work of dealing with the unexpected (Bowker & Star, 1999; Gerson & Star, 1986; Schmidt & Bannon, 1992). Maintenance practices determine a KI's potential for "stability, order, crisis, and breakdown" (Denis & Pontille, 2017, p. 16). Practices include innovating, reimagining, or upgrading aspects of the infrastructure to fight the centrifugal odds of entropy (Denis et al., 2016; Jackson, 2014). Repair work requires improvisation, ingenuity, and flexibility as the work is irreproducible; rarely can it be reduced to static steps (Henke, 1999; Orr, 2016). As the complexity of a KI increases, it becomes "increasingly difficult to define what the 'thing' is that is being maintained…Is it the thing itself, or the negotiated order that surrounds it, or some 'larger' entity?" (Graham & Thrift, 2007, p. 4).

Despite the essential nature of maintenance work in knowledge infrastructures, the labor usually has low status (Morus, 2016; Shapin, 1989). Maintainers are "often among those most marginalized and undervalued in broken systems stemming from established power dynamics" (The Information Maintainers et al., 2019, p. 7). The lower status of maintenance is mirrored in funding structures that value austerity and privatization over maintenance investments, even as infrastructure becomes more fragile with age (Graham & Marvin, 2001).

No one person or institution can maintain a knowledge infrastructure in its entirety. Robust functioning depends upon maintaining individual components and interfaces between them. Pipeline workflows are relatively invisible in comparison to institutional entities such as observatories and data archives. Technical standards such as FITS and value-added services such as CDS and ADS are maintained by their respective institutions and communities. Any component's existence may be obvious to some stakeholders and invisible to others. When a connection between interdependent components breaks down, the fragility of a knowledge infrastructure may suddenly become apparent (Borgman et al., 2016).

In analyzing interview transcripts and notes, we searched for mentions of maintenance and repair activities. We encountered various activities to sustain the technical and physical order of operations, to repair breakdowns that occur, and to avoid future problems. We also looked for other acts of preservation or stewardship of the overall operations. Of particular interest was identifying the specific individuals who perform each kind of maintenance or repair at particular stages of operations. Much of what we label as maintenance in this article falls in the category of "service" in the language of astronomy. Teams large and small have maintenance tasks; everyone does some service to keep the science going.

A scientist who is responsible for several astronomy archives stated the overall challenge thus (2018):

> "[Maintenance] tends not to be addressed, and it's boring. … at some point you have to do it, you have to paint your house, you have to have the roof replaced, it'll happen to every house. … it would be really great if the agencies would say, 'Okay, we need to put money into keeping these going. What are your maintenance needs for the next three years?' …[eventually] you're on a legacy platform that isn't supported anymore, or that has security holes a mile wide…."

## 6.1 Maintaining SDSS Workflows and Data Archives

In astronomy missions of the scale and duration of the Sloan Digital Sky Survey, many stakeholders play many roles. The investigators who write the grant proposals and get proprietary access to data also take on significant management responsibilities. Staff astronomers work on pipelines, data management, instruments, calibration, documentation, help desks, and other tasks. Maintenance and repair pervade these activities. Staff astronomers usually are allocated some time to conduct their own research, which is part of the attraction of being employed on these projects.

SDSS hit many roadblocks in the initial development of the software systems that demanded repairs, "so the software developers were really [the] heroes…called in the middle of the night, night after night…to get these bugs resolved" (SDSS Telescope Engineer, 2012). As one telescope engineer in 2012 explains, maintenance is "constant on the [Sloan Foundation 2.5-Meter Telescope]. We have done a lot of upgrades…, [and] replaced a lot of major components…to make it more efficient." Staff use Arizona's summer monsoon season to invest in major maintenance and repairs when observing is not possible. Senior technical staff design major hardware upgrades such as remapping data acquisition methods. However, as a member of the early SDSS IT staff reported to us in 2012, they may spend a large "amount of [their] professional life being…glorified electrician[s] and air conditioning technician[s]."

As the SDSS phases progressed, the balance of science and service responsibilities shifted. The founding investigators and staff scientists invested many years of development effort before first light. Later SDSS partners had more access to data relative to the service effort required, sometimes causing resentment. As an astronomy professor commented in 2012, "I've put ten years of my career into building this, and these new people …didn't have to build it, … don't have to maintain it and make sure it keeps working, so … have all their time to do science and will get a lot more out of it than we did or we will."

Maintaining the Sloan Digital Sky Survey data, archive, software, and pipelines over the long term is a challenge unto itself. Space-based observatory missions such as Hubble, Chandra, Spitzer, Gaia, and Webb are funded primarily by public sources such as NASA and the European Space Agency, and most include commitments to create and maintain data archives. In contrast, SDSS is a ground-based observatory with primary funding from a private source (the Alfred P. Sloan Foundation). Creating an open data archive was central to the SDSS mission (Brunner et al., 1996; Huang et al., 1995; Thakar et al., 2003). Multiple commitments to house and maintain the data "all have finite lifetimes, and [they] haven't figured out what to do after the lifetime ends" (SDSS scientist, 2012). Among the strategies employed for long-term sustainability was to transfer the SDSS data archive to two research libraries. SDSS investigators continue to seek funding to keep these resources alive (Darch et al., 2020, 2021; Pasquetto et al., 2016).

## 6.2   Maintaining Black Hole Group Workflows and Vida Archives

The Black Hole Group and other teams that partner with the Vida Observatory have vested interests in maintaining their joint and separate infrastructural components. The staff we interviewed and observed at Vida acknowledge the vast scope of their maintenance responsibilities. Observatory staff ensure that instruments continue to be reliable, as measurements tend to drift over time. In analyzing their data from Vida, which continues for months or years after acquiring observations, BHG astronomers are watchful for any systematic errors that might be due to instrument drift or malfunction.

Everyone in the Black Hole Group has a role in maintaining their software tools, including investigators, postdoctoral researchers, graduate students, and research scientists. A postdoctoral researcher explained in 2018 that each member of the group has two roles: "one as a pretty independent scientist…and then [as] support [for] the general, larger project of the group…so everything that has to do with determining orbits around the [black hole], maintaining datasets, [and] taking the observations." They update code, add new functionality, and attempt to root out software rot. Maintenance work necessary to keep these data scientifically valuable includes ensuring they are not corrupted, remain on stable hardware, and are recalibrated to integrate with new data. Service tasks in writing code, maintaining software, and maintaining datasets also are a means for new team members to learn how the science is done. Many of these activities occur on an as-needed basis when something breaks or requires immediate attention to allow continual analyses (Boscoe, 2019).

The temporal nature of BHG science adds a layer of complexity to maintaining their local software and data stores. Because they are continually reintegrating data taken from multiple instruments over very long periods to track orbits around the black hole, they need stable pipelines for comparison. However, research questions change over time, based on new findings, new theories, new instruments, new methods, new investigators, and combinations thereof. New questions require new pipelines or new versions of existing pipelines. The tension between replicability and pursuing new science pervades their software development and maintenance.

Vida Observatory has technical staff responsible for data processing, reduction, maintenance, and transfer to the observatory archive. The Vida Archive is one of several NASA archives, managed by a team responsible for maintaining the technical interfaces, pipelines, and other software. The archive has a ticket system that allows users to flag errors or missing data for archive staff to correct. NASA funding supports baseline maintenance, but our research participants reported difficulty in obtaining support for upgrades to HTML, computer management systems, and other major maintenance and repair.

While the Vida Observatory has institutional responsibility for maintaining instruments and pipelines, as shown in Figure 2, they partner with investigators to identify problems and repair them as necessary. Tensions may arise; for example, in a recent situation where an instrument at Vida was not functioning adequately for the scientific purposes of the BHG, a senior team member devoted an extensive amount of time to debug the problems in collaboration with Vida staff. Some see this effort as uncredited service to the observatory, where others see it as the best means to accelerate their science. All agree that the maintenance work on this instrument needs to be accomplished for the good of the community.

### 6.3 Maintaining Star-Forming Region Survey Data Workflows

The Star-Forming Region Survey relies on the maintenance capabilities of the observatories responsible for the multiple telescopes, instruments, pipelines, and databases from which they collect data. They also rely on a functioning knowledge infrastructure they can use to verify the validity of these datasets. Database documentation, journal articles, standards processes, and other infrastructure components provide bases for judgment.

Tools developed by the Integrative Astronomy Group were part of the data processing pipelines for the Star Forming Region Survey. Some of the IAG tools proved sufficiently useful to be maintained as open-source projects on platforms such as GitHub or BitBucket. At least one of the IAG tools has become a successful commercial product. Because the IAG is a team specializing in open science, pipeline software for the SFRS is probably more readily available and documented than for most astronomy projects. While individual collaborators may still be using parts of the SFRS algorithms, the pipelines are unlikely to be maintained in their entirety 15 or more years after the survey was completed.

The SFRS datasets remain available in the university repository of the project's principal investigator. In this case, the university has agreed to host and curate the datasets indefinitely. Other datasets remain available from the project website. Yet others are available via publishers and authors of papers that used the datasets. As infrastructures evolve, some resources will be migrated, and others will be abandoned.

### 6.4 Theme 3 Discussion

Astronomical artifacts are built to last, or so their designers hope. Observatories such as the Sloan Digital Sky Survey and Vida have multi-decade lifespans, and their data are expected to remain scientifically useful much longer than the physical plant. The need for maintenance and repair of physical structures such as buildings can be predicted in project budgets. Estimating the maintenance required for software and data is a much less exact science (Baker & Karasti, 2018; Karasti et al., 2010; Ribes & Finholt, 2009). Open-source software is particularly complicated to document and maintain. The community is beginning to reckon with the kinds of collaboration required to ensure the stability of essential shared tools such as pipelines (Baker et al., 2016; Darch & Sands, 2017; Geiger et al., 2018; Scroggins et al., 2020).

Our findings reinforce those of Graham and Thrift (2007), that it can be challenging to determine what "thing" is being maintained. All parts of a knowledge infrastructure must be maintained by someone, somehow, at some times, even though no one is in charge of its entirety. Infrastructure components are so deeply intertwined that a breakdown anywhere can cascade through systems. Pipelines are deeply embedded in these interdependent networks. The invisibility of data production work contributes to the difficulty of maintenance and repair; the lack of recognition afforded to those who monitor, fix, and care for data processes and products; and the lack of budget commitment (Denis et al., 2016; Denis & Pontille, 2017; Henke, 1999; Jackson, 2014; Orr, 2016).

## 7 Theme 4: Using and Reusing Data Products

Knowledge production, in the sense explained by Baker and Mayernik (2020), is the primary driver for most scientific endeavors. Scientists collect data as a means to an end, which is to establish their claims by publishing their findings. Scientific knowledge is disseminated through various channels, such as presentations, conference papers, reports, white papers, and other means, but journal articles constitute the core of the intellectual record in the physical, life, and biomedical sciences (Borgman, 2007). Incentives to focus on knowledge production are embedded deeply in the fabric of science, as explained in myriad social, historical, and philosophical studies of science (Latour, 1987; Latour & Woolgar, 1979; Bowker, 2005; Leonelli, 2016; Knorr-Cetina, 1999; Biagioli & Galison, 2003).

The idea that data production should be a central goal of science, or even that data should be produced in ways that make them usable for others, is a recent development. Until the last decade or two, research data in most scientific fields could be controlled indefinitely by the investigators who collected those data. As digital data began to accumulate in large volumes that could be repurposed, funding agencies began to require investigators to release resulting data in a timely manner. Scientific journals followed suit, requiring authors to release data associated with journal articles. The scope of data release policies by funding agencies and journals varies widely across disciplines and countries. Incentives and compliance also appear to vary widely between fields, investigators, and contexts of research. For example, some kinds of data products are amenable to release as discrete, standardized datasets, as in areas of genomics. More commonly, data products are difficult to extricate from their local and temporal contexts without considerable loss of information (Borgman, 2015; Leonelli & Tempini, 2020; Borgman et al., 2010; Bowker, 2005).

Astronomy has been more successful in producing and disseminating datasets in reusable forms than most other sciences. Not least among the many reasons for their success is their extensive knowledge infrastructures that rest on shared agreements for data structures, coordinate systems, and value-added services, as discussed above. Interoperability between datasets, formats, and software tools is essential for integrating datasets from multiple sources, and is a continuing challenge for knowledge infrastructures. The FITS format and tools layered upon these standards are showing their age, raising concerns among astronomers for future interoperability (Mink, 2015; Scroggins & Boscoe, 2020; Thomas et al., 2015; Wang & Xie, 2020). New tools such as Jupyter Notebooks provide scientists the ability to release executable packages of data, pipelines, and workflows. However, these tools appear to improve reusability of data only to the extent that the software and interdependencies continue to be maintained (Wofford et al., 2020).

Data processes, products, incentives, uses, and reuses of data in astronomy vary accordingly, as does the ability to exploit knowledge infrastructures. In this theme, we focus on how data processes contribute to the usability and reusability of data products. Data *reuse* is a broad term in open science, and the one employed in the FAIR principles (Wilkinson et al., 2016). Reuse encompasses access to existing datasets to reproduce a study or replicate findings with new data, to calibrate instruments or results, to combine with other data, or to apply to new scientific problems (Pasquetto et al., 2019; Wallis et al., 2013). Whereas the terms *reproducible* and *replicable* are often used interchangeably, a recent National Academies report establishes important distinctions (Fineberg et al., 2020; Meng, 2020; National Academies of Sciences, Engineering, and Medicine et al., 2019). *Reproducibility*, per the report and as used herein, refers to computational reproducibility using the same input data, methods, and conditions. *Replicability*, in contrast, refers to obtaining consistent results from other studies that address the same scientific question, but using different data.

Because astronomy data are available more widely than in most fields, and are more consistently validated and structured, they are more readily reused for any of these purposes. Hoeppe (2020b) provides a rare case study of how astronomers rely on shared infrastructure to inspect data, challenge results, and to reproduce and replicate empirical findings.

New data releases from major observatories, such as the Gaia release in late 2020, stimulate new science and provide new sources of calibration (European Space Agency, 2020). Scientific journals in astronomy commonly require data to be released at the time of article publication; some now require software citation. Compliance varies, both in the scope of data products and documentation released, and in the degree to which pipeline code used to create those datasets is shared (American Astronomical Society, 2021b; Goodman et al., 2014; Pepe et al., 2014; Shamir et al., 2013; Wofford et al., 2020).

## 7.1 Sloan Digital Sky Survey: Data Use and Reuse

From an archival perspective, it is particularly noteworthy that SDSS data releases are both incremental and cumulative. SDSS scientists reprocess data from prior releases through updated pipelines to improve and correct older observations. By integrating data repeatedly over time, astronomers have access to consistent datasets for reanalysis. By maintaining access to all prior data releases, astronomers also can return to previous states associated with earlier pipelines. When astronomers publish articles using SDSS datasets, they usually cite the specific data release, thus increasing replicability. To ensure continual access to the SDSS datasets by astronomers worldwide, SDSS maintains redundant copies of the data archive at multiple sites.

The community makes extensive use of these data releases. By 2006, the number of hits on the SDSS archive was doubling every year; by 2009 it was estimated that at least 20% of the world's professional astronomers had used Sloan data (Finkbeiner, 2010). A recent estimate is that more than 8000 papers are based on SDSS data, most by non-affiliated scientists (Plant & Hanisch, 2020).

Once available on the SDSS public archive, anyone can download datasets and create copies for local use. Groups that download data from the archive are trusting the quality of data upstream. Research groups often process data further through their private pipelines. Local SDSS datasets, whether downloads of full data releases or small subsets, are "forks" in software terminology. Once "forked" or branched off from the primary dataset for reprocessing, or integrated with other datasets, these resources may take on lives of their own. Local teams are responsible for maintaining and repairing their own pipelines, hardware, and tools. In our studies of SDSS I/II, we found local copies of SDSS datasets all over the world. As astronomers use SDSS data products, they may release reprocessed datasets associated with their publications.

The SDSS data archives are a landmark in open data production both for their scientific contributions and for innovative and unanticipated reuses of those data. Our research participants frequently mentioned cases in which SDSS data were employed to address scientific problems far outside the stated goals of the mission. SDSS became a major site of citizen science with the launch of Galaxy Zoo, at first a simple game-like interface that invited the public to classify galaxies by their spin characteristics. The project was wildly successful in accomplishing its scientific goals, and lead to Zooniverse, a sophisticated citizen science platform that hosts numerous projects (Darch, 2011; Beck et al., 2018; Raddick et al., 2013; Marshall et al., 2015; Borgman et al., 2008; Lintott, 2020). SDSS data products also are being used for music and art installations (Sloan Digital Sky Survey, 2018). SDSS datasets are incorporated in the WorldWide Telescope, a public-facing scientific and educational site developed by Microsoft Research and the American Astronomical Society (American Astronomical Society, 2021a).

## 7.2 Black Hole Group and Vida Observatory: Data Use and Reuse

The Black Hole Group maintains a local repository of all observations they have acquired from the Vida instruments, and from instruments at other observatories, over a period of several decades. BHG team members process, and reprocess, these data through multiple versions of local pipeline software to integrate them with prior observations and to perform new analyses. They describe their first decade of data as "very fragile." While the historical data are of a lesser quality, particularly in terms of image clarity, they remain extremely valuable for following orbits. BHG manages their local data stores in ways that facilitate their own science. They hold individual datasets in multiple states of processing, as observations are combined, recombined, and reassessed based on new findings. These datasets have diverged so extensively from the original observations that BHG members rarely return to their "raw data" in the Vida archive. Boscoe (2019, pp. 71, Figure 4) describes in more detail how BHG reprocesses and reuses their local data stores.

Time-domain astronomy is longitudinal research, akin to long-term data collection in ecology, evolutionary biology, or the humanities. Individual observations are valuable only to the extent that they add information to the cumulative record. A scholar in the humanities referred to her longitudinal corpora as her "dowry" to leverage in finding academic positions and in attracting collaborators (Borgman, 2015). Scholars in the sciences similarly use these comprehensive corpora as "barter" to attract collaborators and students (Hilgartner & Brandt-Rauf, 1994). In the environmental sciences, we found that scholars engaged in longitudinal research were least willing to release their corpora (Borgman et al., 2010). The Long Term Ecological Research Centers were established as means to collect and disseminate longitudinal data products and to develop new data practices (Kaplan et al., 2021; Karasti et al., 2010; Waide & Kingsland, 2021). Similarly, the BHG team has few incentives to release their entire corpora. Like other scientists conducting longitudinal research, they release datasets associated with individual articles, at the time of publication.

Funding for the Black Hole Group consists largely of overlapping research grants. Individual grant projects advance their scientific methods by developing software tools, contributing to instrument design, collecting new observations, and writing papers. BHG is not funded as a scientific mission with dedicated archivists and other support staff. All team members participate in building and maintaining their local data stores, which consist of many generations of observations, software pipelines for photometry, astrometry, and spectroscopy; other code and algorithms to clean, reduce, validate, and compare observations; tutorials and "cookbooks" for observing and processing data; and miscellaneous forms of documentation. These scientists produce research-grade software on which they train new students and staff as they join the group. They use and reuse data products continuously.

The Black Hole Group is among the research teams who were observing at Vida long before a public archive existed. They had little choice but to develop their own pipeline software, data management practices, and local data stores (Boscoe, 2019). Vida has maintained backup copies of all data acquired since first light in the mid-1990s, but the mission did not include a public archive in its initial funding. About a decade after first light, NASA funded the Vida Observatory Archive. Vida now imposes a default 18-month proprietary period over observations, unless investigators negotiate a longer control period. BHG, along with other investigators who observe at Vida, benefit from the observatory's knowledge infrastructure investments in the data archive, thus assuring long-term access to datasets collected with their telescopes and instruments.

### 7.3 Star Forming Region Survey: Data Use and Reuse

Data products from the Star-Forming Region Survey continue to be available to the astronomy community long after the collaboration ended. The SFRS team members, as pioneers of open astronomical data, wanted to provide data products that were reusable in practice. They facilitate reuse by providing multiple entry points into the data. Entry points include a browsable data table and an interactive tool that shows data coverage by region. The team also clarified that the data are non-proprietary and can be cited using a data paper to promote reuse and with appropriate attribution. The published datasets and associated data papers from the survey have received more than 900 citations in the astronomy literature as of this writing.

The SFRS posted data products acquired from public sources, both in original forms and in their re-reduced forms. Per the principal investigator (2020 followup interview), the SFRS data are now easier to find than the original source data, given numerous moves and redirects in the two decades or so since the space telescope mission acquired those observations. Published datasets remain available at the university's open data repository and on project web page, thus facilitating access for the community.

### 7.4 Theme 4 Discussion

Our participants rarely mentioned terms such as reproducible or replicable research; they focus more generally on the ability to reuse their own and others' datasets. Versions of astronomy datasets proliferate around the world as data are used and reused by the community. Software does not always follow the data; rarely did our participants indicate that data users had access to the code employed to process observations into scientific data products. Our findings reveal the iterative methods and multiple tools that comprise astronomical pipelines. Astronomy data products take on lives of their own, requiring each subsequent user to trust upstream processing. While all three of these projects are concerned with creating, maintaining, using, and reusing data products from astronomical observations, their scientific approaches and archival concerns diverge.

The Sloan Digital Sky Survey is a four-decade mission whose data serve a wide array of scientific problems, are open to the community, and are reused intensively by Sloan investigators and by astronomers unaffiliated with the project. SDSS continues to provide access to all sixteen data releases. Beyond these official data releases are a diaspora of copies, which we found distributed around the world. Individual astronomy teams download SDSS data products. Some use these datasets for a publication or two; others maintain SDSS subsets locally for further reprocessing and research.

The Black Hole Group is in its third decade of studying the physics of black holes and their environments; these are temporal experiments that require new data collection on a regular basis. They maintain all of their data products and software locally for reintegration and reuse. Somewhat analogous to SDSS, they maintain incremental datasets to ensure integrity of their observational record, while moving cumulative datasets forward as they reprocess data to reflect new methods and pipelines. They release processed datasets associated with individual journal articles, but only BHG team members have access to the full corpora, reflecting common practice in the field. As a rare longitudinal dataset in astronomy, the corpora is their dowry to attract students and collaborators.

The Vida Observatory is in its third decade of access to astronomers for data collection, whereas the Vida Observatory Archive is in its second decade of providing public access to data collected by those astronomers. Together, observatory and archive staff manage the data processing pipelines to merge observations from multiple instruments on the telescopes into a common format. These data products are all incremental, comprised of disparate observations of disparate targets, under disparate instrument settings, with varying methodologies, for a vast array of scientific goals, taken by individual investigators. Other scientists can obtain these data products after proprietary periods are complete. Archival data products are valuable records of measurements of specific targets at specific times, but they are relatively "raw data." Vida, like most data archives, is not in the business of producing comprehensive data releases accompanied by extensive documentation. That is the role of long-term surveys such as SDSS.

The Star-Forming Region Survey was a short-term entity and the Integrative Astronomy Group is an informal collaboration. However, their data products may remain useful indefinitely, with or without the processing software used to create them.

**Figure 5: Three-Stream Model of Knowledge and Data Production.** Adapted from two-stream model of Baker and Mayernik (2020) in "Disentangling knowledge production and data production" licensed under [CC BY 4.0](CC BY 4.0)

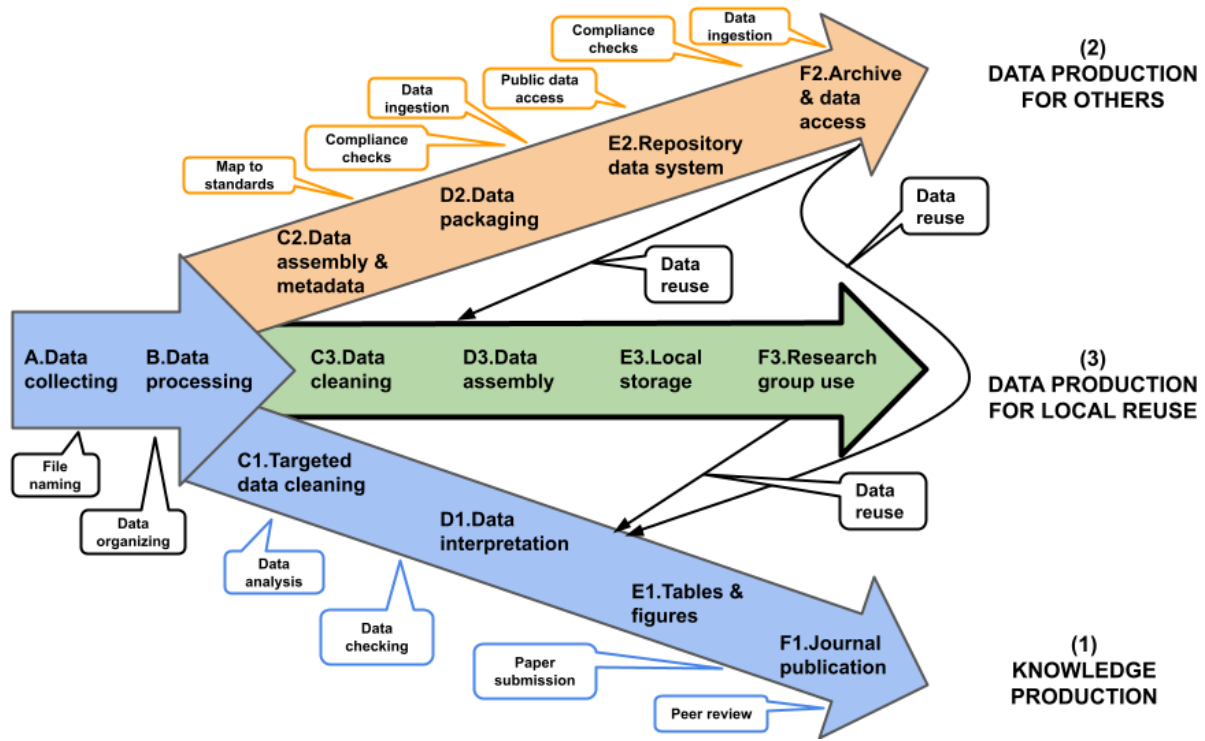

As shown in Figure 5, we extend the Baker and Mayernik (2020) dichotomy of knowledge production (blue arrow (1) at bottom) vs. data production for use by others (peach arrow (2) at top), by adding a third goal (green arrow (3) in center) of data production for local reuse. All scientific projects have knowledge production goals; they vary in the degree to which they also invest in curating data for others in the community. The Sloan Digital Sky Survey has the strongest commitment to open data production of the three projects studied and was a pioneer in these endeavors. The SDSS scientific data archiving team performs the curation activities along the peach arrow (2), which enables both SDSS investigators and later reusers to engage in knowledge production from those data (blue arrow 1).

The Black Hole Group, while unusual in conducting time-domain astronomy that requires such a long-term dataset, is more typical of scientific investigations in that they focus on maintaining their data stores for use by their own team. BHG typifies our third stream (green arrow (3)). Team members clean, assemble, and write new code for their cumulative data store (activities in (3)), which enables partners – past, present, and future – to produce new knowledge from those data products (arrow 1). BHG uses their data store to attract and train new students and collaborators, transferring tacit knowledge from one generation to the next. The Vida Observatory Archive performs the activities in the peach arrow (2) that enable others to use BHG raw data after proprietary periods end. Other scientists also can reuse processed datasets attached to BHG journal articles (arrow 1).

The partnership of the Integrated Astronomy Group and the Star-Forming Region Survey was an early example of an open science project with dual goals of knowledge and data production for others (arrows 1 and 2). They released their data immediately, even before publication of their articles. They have kept the project website available with downloadable datasets and deposited their datasets in a university data archive. Both of these accomplishments are unusual and beyond the resources of most scientists. However, data access differs between SDSS and SFRS in ways that are important for long-term scientific reuse. SDSS is an active data archive with curatorial staff, a help desk, ticket system for errors, and the means to assist scientists in searching for data, all of which contribute to sustainability. SFRS data products are preserved as static datasets in a generic university repository. Digital bits are sustained in the repository and on the website, but scientific assistance is limited to contacting investigators associated with work done two decades earlier.

Datasets attached to papers facilitate replicability, but rarely cumulate to become coherent archives (Borgman, 2015). Comparing these three projects reveals how differences in commitments to knowledge production and data production determine what data become available to the community (Baker & Mayernik, 2020). Science progresses one project and one paper at a time. Data production facilitates science by creating shared community resources.

# 8   Conclusions

Astronomy is a data-intensive research domain that depends upon large scale, international, and collaborative knowledge infrastructures. They have invested heavily in observatories, instrumentation, technical standards, data archives, and value-added services. The least visible components of these knowledge infrastructures are the software tools used to process observations into scientifically useful data products. It has taken us more than a decade of studying scientific practice in this field to tease out the subtle and profound ways in which astronomy research methods are enacted in algorithms. These algorithms incorporate astronomers' recipes for identifying scientific targets precisely, for extracting signals from noise, and for combining and comparing observations. Software pipelines are crucial, but fragile and often brittle, components of astronomical knowledge infrastructures.

Scientific software, especially research-grade software created for local purposes, is notoriously difficult to maintain. Due to the elaborate interdependencies among software tools, any change or upgrade may fix one problem while introducing problems or incompatibilities elsewhere. Pipelines evolve with science and individual scientists, splitting and merging and branching over time. Data products similarly evolve, as the same observations from a telescopic instrument may be processed through multiple pipelines by observatory staff, investigators, and archives, and then reprocessed many more times as these data become evidence of other phenomena, or serve other scientific questions. Some pipelines are published and maintained by the community, but many remain local and proprietary. Without explicit commitments for curation, most code ceases to function fairly quickly. Fragility is the norm; durability is the exception.

Knowledge production is the driving force of scientific projects. Processing observations into forms that become useful for the investigators is itself a major endeavor that requires extensive software development, as demonstrated by the three projects studied here. Substantial additional processing is necessary to package and curate these data products in ways that make them reusable by a broader community. As Baker and Mayernik (2020) explain, orientation toward either knowledge production or public data release strongly influences project organization and funding; these distinctions rarely are explicit commitments. Such choices become apparent when tensions arise over relative investments in maintenance, archiving, software development, curation and metadata, publishing practices, data release, open science, and related matters. When infrastructures break down, whether archives or pipelines, implicit assumptions may suddenly become explicit decisions.

Astronomers commonly reuse reduced data without access to the original pipeline software, as exhibited by the extensive reuse of SDSS and Star-Forming Region Survey data products. Sky surveys are a specific kind of astronomy research that collects consistent observations of specific phenomena in ways intended for comparisons by others. These datasets are released with extensive documentation. Observatory archives such as Vida also document data products they release for public use. However, these are heterogenous datasets that reflect the diversity of scientific purposes served by these observatories. Another important distinction is that most space-based observatories, such as Hubble, Chandra, and Gaia, maintain data archives for the astronomy community, whereas few ground-based observatories have dedicated funding for data archives. Vida is an exception. SDSS, as a ground-based survey operating on grants for individual phases, continues to seek funding to maintain their data archive.

Long-term scientific investigations, such as that conducted by the Black Hole Group, necessarily focus on knowledge production. The team has vested interests in keeping these data "alive" for local reuse (Boscoe, 2019). Many hands are involved, as data and software evolve over the course of many years, many projects, and many technologies. All of their data remain available to the team, albeit in multiple states of processing. The older the data, the greater the distance from the technologies used to capture and store those data. Accordingly, data are reprocessed many times over their lifespan as they are integrated with other sources. Maintaining the viability of all software versions employed over several decades is nigh unto impossible. As their data products accumulate, sustainability becomes an ever larger challenge, and one for which it is difficult to acquire research funding. Grants support each new round of observations and software development, but not the continual reinvestment in the local data stores that are essential to their science. Our astronomy participants frequently mentioned their difficulties in obtaining support for data archiving from grant funds or from their universities. The lack of mechanisms to support data archiving remains a grand challenge for open science.

Data products from the Star-Forming Region Survey remain available more than 15 years on, which is unusual. They remain available because the principal investigator's university has agreed to do so, not out of continuing support by any funding agency.

As requested by astronomers who participate in our studies, we conclude with some knowledge infrastructure recommendations for scientific stakeholders. One recommendation is to make explicit distinctions between knowledge production and open data release in grant funding, scientific policy, and institutional practice. Scientists are funded to produce knowledge in the form of publications, which may also include datasets. Rarely are they funded to be data archivists. If a scientific team is to create data resources for public use, and to sustain access to those resources, additional funds and different skill sets are required.

A related recommendation is to invest in data sources that have long-term value. Most space-based missions are funded with data archives; these are heavily exploited by their communities. Ground-based missions such as Sloan and Vida have struggled to obtain comparable commitments. Our participants view space-based and ground-based data as having comparable scientific value; the differential investments appear to be due to community practice and agency policies.

Software sustainability remains a massive challenge throughout science, education, government, and industry. Astronomers, like other scientists, are publishing more of their software and contributing open source software on platforms such as GitHub and BitBucket. Collaborative commitments to maintain useful pipelines, combined with good documentation and established technical standards, are more feasible approaches to sustaining access to data than simply publishing pipeline code. As the field grows to rely ever more heavily on open source software, determining collective responsibility for maintaining infrastructure will be challenging.

Lastly, we encourage more investment in the staff who sustain the knowledge infrastructures that make science possible. These scientists, technologists, archivists, and other specialists are the unsung heroines and heroes of knowledge and data production. The community needs to develop robust career paths for the personnel who keep fragile systems from breaking down, and who repair them when they do. Long after spacecraft become space junk, the data remain a valuable scientific investment. No matter how advanced these technologies, sustainability still requires humans in the loop.

## Disclosure Statement

Research reported here was supported in part by grants from the National Science Foundation (#1145888, C.L. Borgman, PI; S. Traweek, Co-PI, and #0830976), and the Alfred P. Sloan Foundation (#20113194, C.L. Borgman, PI; S. Traweek, Co-PI, and #201514001, C.L. Borgman, PI), and by the Harvard-Smithsonian Center for Astrophysics for hosting C.L. Borgman as a Visiting Scientist in 2018. Morgan F. Wofford completed most of her work on this article in her role as Data Scientist for the UCLA Center for Knowledge Infrastructures with funding from the Alfred P. Sloan Foundation Digital Information Technology Program, directed by Joshua M. Greenberg.

## Acknowledgements
We thank Bernadette M. Boscoe, Peter T. Darch, Margaret G. Kivelson, and Michael J. Scroggins for commenting on early drafts. Karen S. Baker and Matthew S. Mayernik provided comments on multiple drafts and kindly gave permssion to reproduce key parts of their model in our Figure 5. The Editor-in-Chief, Associate Editor, and two reviewers of HDSR provided exceptionally constructive guidance. We thank members of the Sloan Digital Sky Survey, Black Hole Group, and Integrative Astronomy Group for thoughtful corrections and improvements to this article. Astronomy interviews and notes analyzed for this article were collected from 2008 through 2021 by Christine L. Borgman, Bernadette M. Boscoe, Peter T. Darch, David S. Fearon Jr., Ashley E. Sands, Michael J. Scroggins, Sharon Traweek, and Laura A. Wynholds, with additional data analysis by Milena Golshan. Most of all, we are grateful to the many members of the astronomy community who granted us interviews, access to their laboratories and offices, and provided rare or unpublished documents over the last 15 years.

Borgman & Wofford, From Data Processes to Data Products, Authors Accepted version, June 2021, Page 35 of 37Pasquetto, I. V., Sands, A. E., Darch, P. T., & Borgman, C. L. (2016). Open Data in Scientific Settings: From Policy to Practice. *Proceedings of the 2016 CHI Conference on Human Factors in Computing Systems*, 1585–1596. https://doi.org/10.1145/2858036.2858543

Pepe, A., Goodman, A., Muench, A., Crosas, M., & Erdmann, C. (2014). How Do Astronomers Share Data? Reliability and Persistence of Datasets Linked in AAS Publications and a Qualitative Study of Data Practices among US Astronomers. *PLOS ONE*, *9*(8), e104798. https://doi.org/10.1371/journal.pone.0104798

Plant, A. L., & Hanisch, R. J. (2020). Reproducibility in Science: A Metrology Perspective. *Harvard Data Science Review*, *2*(4). https://doi.org/10.1162/99608f92.eb6ddee4

Raddick, M. J., Bracey, G., Gay, P. L., Lintott, C. J., Cardamone, C., Murray, P., Schawinski, K., Szalay, A. S., & Vandenberg, J. (2013). Galaxy Zoo: Motivations of citizen scientists. *Astronomy Education Review*, *12*(1). https://doi.org/10.3847/AER2011021

Ribes, D., & Finholt, T. (2009). The Long Now of Technology Infrastructure: Articulating Tensions in Development. *Journal of the Association for Information Systems*, *10*(5). https://doi.org/10.17705/1jais.00199

Sands, A. E. (2017). *Managing Astronomy Research Data: Data Practices in the Sloan Digital Sky Survey and Large Synoptic Survey Telescope Projects* [Ph.D. Dissertation, UCLA]. http://escholarship.org/uc/item/80p1w0pm

Schmidt, K., & Bannon, L. (1992). Taking CSCW seriously: Supporting articulation work. *Computer Supported Cooperative Work (CSCW)*, *1*(1), 7–40. https://doi.org/10.1007/BF00752449

Scroggins, M. J., & Boscoe, B. M. (2020). Once FITS, Always FITS? Astronomical Infrastructure in Transition. *IEEE Annals of the History of Computing*, *42*(2), 42–54. https://doi.org/10.1109/MAHC.2020.2986745

Scroggins, M. J., & Pasquetto, I. V. (2020). Labor Out of Place: On the Varieties and Valences of (In)visible Labor in Data-Intensive Science. *Engaging Science, Technology, and Society*, *6*(0), 111–132. https://doi.org/10.17351/ests2020.341

Scroggins, M. J., Pasquetto, I. V., Geiger, R. S., Boscoe, B. M., Darch, P. T., Cabasse-Mazel, C., Thompson, C., Golshan, M. S., & Borgman, C. L. (2020). Thorny problems in data (-intensive) science. *Communications of the ACM*, *63*(8), 30–32. https://doi.org/10.1145/3408047

SDSS Collaboration. (2014, November 10). *The SDSS Science Legacy*. Sloan Digital Sky Survey (Classic). http://classic.sdss.org/signature.html

SDSS Collaboration. (2015, January 3). *Sloan Digital Sky Surveys*. SDSS. http://www.sdss.org/sdss-surveys/

SDSS Collaboration. (2020a). *SDSS-V: Pioneering Panoptic Spectroscopy | SDSS*. https://www.sdss.org/future/

SDSS Collaboration. (2020b). *Sloan Digital Sky Surveys | SDSS*. https://www.sdss.org/surveys/

SDSS Collaboration. (2020c). *Telescopes and Instruments | SDSS*. https://www.sdss.org/instruments/

Shamir, L., Wallin, J. F., Allen, A., Berriman, B., Teuben, P., Nemiroff, R. J., Mink, J. D., Hanisch, R. J., & DuPrie, K. (2013). Practices in source code sharing in astrophysics. *Astronomy and Computing*, *1*, 54–58. https://doi.org/10.1016/j.ascom.2013.04.001

Shapin, S. (1989). The Invisible Technician. *American Scientist*, *77*(6), 554–563. http://www.jstor.org/stable/27856006

Sloan Digital Sky Survey. (2018). *The Sloan Digital Sky Survey names its first artist in residence*. https://www.sdss.org/press-releases/artist-in-residence/